\documentclass[12pt,
showpacs,preprintnumbers,superscriptaddress,nofootinbib]{revtex4}
\usepackage{epsfig} \usepackage{graphicx}

\begin{document}

\def\simge{\hspace*{0.2em}\raisebox{0.5ex}{$>$}
     \hspace{-0.8em}\raisebox{-0.3em}{$\sim$}\hspace*{0.2em}}
     \def\simle{\hspace*{0.2em}\raisebox{0.5ex}{$<$}
     \hspace{-0.8em}\raisebox{-0.3em}{$\sim$}\hspace*{0.2em}}
     \def\bra#1{{\langle#1\vert}} \def\ket#1{{\vert#1\rangle}}
     \def\coeff#1#2{{\scriptstyle{#1\over #2}}}
     \def\undertext#1{{$\underline{\hbox{#1}}$}}
     \def\hcal#1{{\hbox{\cal #1}}} \def\sst#1{{\scriptscriptstyle #1}}
     \def\eexp#1{{\hbox{e}^{#1}}} \def\rbra#1{{\langle #1
     \vert\!\vert}} \def\rket#1{{\vert\!\vert #1\rangle}} \def\lsim{{
     <\atop\sim}} \def\gsim{{ >\atop\sim}}
\def\nubar{{\bar\nu}} \def\psibar{{\bar\psi}} \def\Gmu{{G_\mu}}
\def\alr{{A_\sst{LR}}} \def\wpv{{W^\sst{PV}}} \def\evec{{\vec e}}
\def\notq{{\not\! q}} \def\notk{{\not\! k}} \def\notp{{\not\! p}}
\def\notpp{{\not\! p'}} \def\notder{{\not\! \partial}}
\def\notcder{{\not\!\! D}} \def\notA{{\not\!\! A}} \def\notv{{\not\!\!
v}} \def\Jem{{J_\mu^{em}}} \def\Jana{{J_{\mu 5}^{anapole}}}
\def\nue{{\nu_e}} \def\mn{{m_\sst{N}}} \def\mns{{m^2_\sst{N}}}
\def\me{{m_e}} \def\mes{{m^2_e}} \def\mq{{m_q}} \def\mqs{{m_q^2}}
\def\mz{{M_\sst{Z}}} \def\mzs{{M^2_\sst{Z}}} \def\mws{{M^2_\sst{W}}}
\def\ubar{{\bar u}} \def\dbar{{\bar d}} \def\sbar{{\bar s}}
\def\qbar{{\bar q}} \def\sstw{{\sin^2\theta_\sst{W}}}
\def\gv{{g_\sst{V}}} \def\ga{{g_\sst{A}}} \def\pv{{\vec p}}
\def\pvs{{{\vec p}^{\>2}}} \def\ppv{{{\vec p}^{\>\prime}}}
\def\ppvs{{{\vec p}^{\>\prime\>2}}} \def\qv{{\vec q}} \def\qvs{{{\vec
q}^{\>2}}} \def\xv{{\vec x}} \def\xpv{{{\vec x}^{\>\prime}}}
\def\yv{{\vec y}} \def\tauv{{\vec\tau}} \def\sigv{{\vec\sigma}}
\def\sst#1{{\scriptscriptstyle #1}} \def\gpnn{{g_{\sst{NN}\pi}}}
\def\grnn{{g_{\sst{NN}\rho}}} \def\gnnm{{g_\sst{NNM}}}
\def\hnnm{{h_\sst{NNM}}}

\def\xivz{{\xi_\sst{V}^{(0)}}} \def\xivt{{\xi_\sst{V}^{(3)}}}
\def\xive{{\xi_\sst{V}^{(8)}}} \def\xiaz{{\xi_\sst{A}^{(0)}}}
\def\xiat{{\xi_\sst{A}^{(3)}}} \def\xiae{{\xi_\sst{A}^{(8)}}}
\def\xivtez{{\xi_\sst{V}^{T=0}}} \def\xivteo{{\xi_\sst{V}^{T=1}}}
\def\xiatez{{\xi_\sst{A}^{T=0}}} \def\xiateo{{\xi_\sst{A}^{T=1}}}
\def\xiva{{\xi_\sst{V,A}}}

\def\rvz{{R_\sst{V}^{(0)}}} \def\rvt{{R_\sst{V}^{(3)}}}
\def\rve{{R_\sst{V}^{(8)}}} \def\raz{{R_\sst{A}^{(0)}}}
\def\rat{{R_\sst{A}^{(3)}}} \def\rae{{R_\sst{A}^{(8)}}}
\def\rvtez{{R_\sst{V}^{T=0}}} \def\rvteo{{R_\sst{V}^{T=1}}}
\def\ratez{{R_\sst{A}^{T=0}}} \def\rateo{{R_\sst{A}^{T=1}}}

\def\mro{{m_\rho}} \def\mks{{m_\sst{K}^2}} \def\mpi{{m_\pi}}
\def\mpis{{m_\pi^2}} \def\mom{{m_\omega}} \def\mphi{{m_\phi}}
\def\Qhat{{\hat Q}}

\def\FOS{{F_1^{(s)}}} \def\FTS{{F_2^{(s)}}}
\def\GAS{{G_\sst{A}^{(s)}}} \def\GES{{G_\sst{E}^{(s)}}}
\def\GMS{{G_\sst{M}^{(s)}}} \def\GATEZ{{G_\sst{A}^{\sst{T}=0}}}
\def\GATEO{{G_\sst{A}^{\sst{T}=1}}} \def\mdax{{M_\sst{A}}}
\def\mustr{{\mu_s}} \def\rsstr{{r^2_s}} \def\rhostr{{\rho_s}}
\def\GEG{{G_\sst{E}^\gamma}} \def\GEZ{{G_\sst{E}^\sst{Z}}}
\def\GMG{{G_\sst{M}^\gamma}} \def\GMZ{{G_\sst{M}^\sst{Z}}}
\def\GEn{{G_\sst{E}^n}} \def\GEp{{G_\sst{E}^p}}
\def\GMn{{G_\sst{M}^n}} \def\GMp{{G_\sst{M}^p}}
\def\GAp{{G_\sst{A}^p}} \def\GAn{{G_\sst{A}^n}} \def\GA{{G_\sst{A}}}
\def\GETEZ{{G_\sst{E}^{\sst{T}=0}}}
\def\GETEO{{G_\sst{E}^{\sst{T}=1}}}
\def\GMTEZ{{G_\sst{M}^{\sst{T}=0}}}
\def\GMTEO{{G_\sst{M}^{\sst{T}=1}}}
\def\lamd{{\lambda_\sst{D}^\sst{V}}} \def\lamn{{\lambda_n}}
\def\lams{{\lambda_\sst{E}^{(s)}}} \def\bvz{{\beta_\sst{V}^0}}
\def\bvo{{\beta_\sst{V}^1}} \def\Gdip{{G_\sst{D}^\sst{V}}}
\def\GdipA{{G_\sst{D}^\sst{A}}} \def\fks{{F_\sst{K}^{(s)}}}
\def\FIS{{F_i^{(s)}}} \def\fpi{{F_\pi}} \def\fk{{F_\sst{K}}}

\def\RAp{{R_\sst{A}^p}} \def\RAn{{R_\sst{A}^n}}
\def\RVp{{R_\sst{V}^p}} \def\RVn{{R_\sst{V}^n}}
\def\rva{{R_\sst{V,A}}} \def\xbb{{x_B}}

\def\PR#1{{{\em Phys. Rev.} {\bf #1} }} \def\PRC#1{{{\em Phys. Rev.}
{\bf C#1} }} \def\PRD#1{{{\em Phys. Rev.} {\bf D#1} }}
\def\PRL#1{{{\em Phys. Rev. Lett.} {\bf #1} }} \def\NPA#1{{{\em
Nucl. Phys.} {\bf A#1} }} \def\NPB#1{{{\em Nucl. Phys.} {\bf B#1} }}
\def\AoP#1{{{\em Ann. of Phys.} {\bf #1} }} \def\PRp#1{{{\em
Phys. Reports} {\bf #1} }} \def\PLB#1{{{\em Phys. Lett.} {\bf B#1} }}
\def\ZPA#1{{{\em Z. f\"ur Phys.} {\bf A#1} }} \def\ZPC#1{{{\em
Z. f\"ur Phys.} {\bf C#1} }} \def\etal{{{\em et al.}}}

\def\delalr{{{delta\alr\over\alr}}} \def\pbar{{\bar{p}}}
\def\lamchi{{\Lambda_\chi}} \newcommand{\amulbl}{a_\mu^{\sst{LL}}}
\def\rnu{{R_\nu}} \def\rnubar{{R_{\bar\nu}}}
\def\sinhat{\sin^2\hat\theta_W}

\def\alred{{A_{\sst LR}^{eD,\ \rm DIS}}}

\newcommand{\slashq}{\not{\hbox{\kern-3pt $q$}}}
\def\stilde{\widetilde} \newcommand{\beqa}{\begin{eqnarray}}
\newcommand{\eeqa}{\end{eqnarray}} \newcommand{\beq}{\begin{equation}}
\newcommand{\eeq}{\end{equation}}



\preprint{CALT-68-2446} \preprint{MAP-291} \preprint{hep-ph/0307270}

\title{Supersymmetric Effects in Parity-Violating Deep Inelastic
Electron-Nucleus Scattering}


\author{A. Kurylov} \affiliation{ California Institute of Technology,
Pasadena, CA 91125\ USA}

\author{M.J. Ramsey-Musolf} \affiliation{ California Institute of
Technology, Pasadena, CA 91125\ USA} \affiliation{ Department of
Physics, University of Connecticut, Storrs, CT 06269\ USA}

\author{S.~Su} \affiliation{ California Institute of Technology,
Pasadena, CA 91125\ USA}



\begin{abstract}
We compute the supersymmetric (SUSY) corrections to the
parity-violating, deep inelastic electron-deuteron asymmetry. Working
with the Minimal Supersymmetric Standard Model (MSSM) we consider two
cases: R parity conserving and R parity-violating. Under these
scenarios, we compare the SUSY effects with those entering other
parity-violating observables.  For both cases of the MSSM, we find
that the magnitude of the SUSY corrections can be as large as $\sim
1\%$ and that they are strongly correlated with the effects on other
parity-violating observables. A comparison of various low-energy
parity-violating observables thus provides a potentially interesting
probe of SUSY.

\end{abstract}


\maketitle

\vspace{0.3cm}

\pagenumbering{arabic}


\section{Introduction}
\label{intro}

The study of parity-violation (PV) in nuclei and atoms is playing an
important role in the search for physics beyond the Standard Model
(SM) \cite{mrm99}. Historically, the first such study in the neutral
current sector involved a measurement of the PV \lq\lq left-right"
asymmetry, $\alr$, for deep inelastic electron-deuteron
scattering\cite{pre78}. That experiment, which established the
validity of the SM for weak neutral currents, was followed by several
studies of parity-violation in atoms\cite{ben99} as well as in both
quasi-elastic\cite{hei89} and elastic electron-nucleus
scattering\cite{sou90}. Recently, the first results have been reported
for a measurement of $\alr$ for $e^-e^-$ scattering at the Stanford
Linear Accelerator Center (SLAC)\cite{slac,sou03}, and a proposal to
measure $\alr$ in elastic $ep$ scattering has been approved at the
Jefferson Laboratory (JLab)\cite{car02}.  The precision expected for
the latter two measurements allows one both to test the
$q^2$-evolution of the weak mixing angle as well as to probe for new
physics. In light of these developments, ideas have been generated for
a new generation of PV DIS measurements with deuterium targets at both
SLAC\cite{slacloi} and JLab\cite{pcdr}.

In this note, we analyze the prospective implications of new PV DIS
measurements for supersymmetric extensions of the Standard
Model. Supersymmetry (SUSY) remains one of the most strongly-motivated
new physics scenarios, and one hopes that future collider measurements
will provide the first direct evidence for its existence. At present,
the phenomenological evidence for SUSY, derived from precision
electroweak measurements, is sparse. Although the new measurements of
the muon anomalous magnetic moment\cite{muon} generated considerable
excitement in the SUSY community, the initial indications of a
deviation from the SM have been superseded by subsequent developments
in SM theory. In the charged current sector, the long-standing
deviation from unitarity of the CKM matrix -- taken in conjunction with
measurements of the $W$-mass and muon anomaly -- could signal the
insufficiency of standard SUSY-breaking models\cite{kurylov02}, though
recent experimental and theoretical developments\cite{sher,bij03}
involving kaon leptonic decays have clouded the situation somewhat. The
implications of precision neutral current experiments are mixed as
well. Fits to Z-pole observables, performed using different
assumptions about the mechanism of SUSY-breaking mediation, have lead
to varying conclusions about whether or not the data favor the
presence of SUSY\cite{hag94,erl98}. At much lower energy scales, the
results of cesium atomic PV are consistent with both the SM \cite{apv}
as well as its minimal SUSY extensions\cite{susypv}. In contrast, the
substantial deviation from the SM in deep-inelastic neutrino-nucleus
scattering reported by the NuTeV collaboration \cite{nutev02} cannot
be explained using any of the most common SUSY
scenarios\cite{susynutev}. In principle, the new PV $ee$ and $ep$
measurements will help clarify the low-energy neutral current
situation.

In this context, one would like to know what new insight -- if any --
a precise measurement of the PV DIS $eD$ asymmetry might provide. To
that end, we compute the SUSY electroweak radiative corrections to
this asymmetry using the Minimal Supersymmetric Standard Model
(MSSM). We also analyze the possible effects of SUSY interactions that
violate R parity. In doing so, we compare the effects of either
scenario on the DIS asymmetry with the corresponding effects in other
low-energy neutral current experiments. As we noted in our previous
studies of PV $ee$ and $ep$ scattering\cite{susypv}, we find that
there exist notable correlations involving the SUSY effects on
different observables. We also work out the level of precision needed
for a PV DIS measurement in order for it to be a significant probe of
SUSY.

The PV asymmetry for deep inelastic $eD$ scattering, $\alred$, has the
simple form \begin{equation}
\label{eq:pvasym}
\alred=\frac{\sigma_R-\sigma_L}{\sigma_R+\sigma_L}={G_\mu Q^2\over
4\sqrt{2}\pi\alpha}{9\over 5}\left\{ {\tilde a}_1+ {\tilde
a}_2\left[{1-(1-y)^2\over 1+(1-y)^2}\right]\right\}\ \ \ ,
\end{equation}
where
\begin{eqnarray}
\label{eq:atilde}
{\tilde a}_1 & = & \frac{2}{3}(2 C_{1u} -C_{1d})\\ {\tilde a}_2 & = &
\frac{2}{3}(2 C_{2u} -C_{2d})\ \ \ \nonumber
\end{eqnarray}
and $y\in [0,1]$ is the fractional energy transfer to the target (in the lab
frame)\footnote{For simplicity, we have neglected a correction to the 
denominator of the $y$-dependent term in Eq. (\ref{eq:pvasym}) arising 
from longitudinal gauge boson contributions\cite{Blum89}.}.  
The quantities $C_{iq}$ parameterize the low-energy, PV
electron-quark interaction
\begin{equation}
{\cal L}^{eq}_{\rm PV} = {G_\mu\over \sqrt{2}}\sum_q\ \left[ C_{1q}
{\bar e}\gamma^\mu\gamma_5 e {\bar q}\gamma_ \mu q \ + \ C_{2q} {\bar
e}\gamma^\mu e {\bar q} \gamma_\mu\gamma_5 q\right] \ \ \ .
\end{equation}
Note that in writing down Eqs.~(\ref{eq:pvasym},{\ref{eq:atilde}), we
have neglected target mass and higher-twist corrections as well as
contributions from sea quarks\footnote{While the magnitude and
theoretical reliability for these corrections are important for the
interpretation of $\alred$, we do not consider them in detail here.}.

The presence of SUSY interactions will modify the $C_{iq}$ from their
SM values. The $C_{iq}$ are conveniently computed using the
expressions
\begin{eqnarray}
C_{1q} & = & 2{\rho}_{PV} I_3^e(I_3^q-2 Q_q { \kappa}_{PV}
{\hat s}^2)-\frac{1}{2}{\hat \lambda}_{1q} \\ C_{2q} & = &
2{\rho}_{PV} I_3^q(I_3^e-2 Q_e{\kappa}_{PV} {\hat
s}^2)-\frac{1}{2}{\hat \lambda}_{2q}\ \ \ ,
\end{eqnarray}
where $I_3^f$ and $Q_f$ are, respectively, the third component of weak
isospin and electric charge of fermion $f$, ${\hat
s}=\sin{\hat\theta_\sst{W}}$ is the sine of the mixing angle in the
$\overline{\rm MS}$ scheme, and ${\rho}_{PV}=1={\kappa}_{PV}$
and $\hat\lambda_{iq}=0$ at tree-level in the SM. Electroweak
radiative corrections modify the latter quantities from their
tree-level values: $\rho_{PV}=1+\delta{\rho}_{PV}$,
${\kappa}_{PV} = 1+\delta{\kappa}_{PV}$, with $\delta\rho_{PV}$,
$\delta{\kappa}_{PV}$, and $\hat\lambda_{iq}$ being of order
${\alpha}$.

\begin{figure}
\begin{center}
\resizebox{16. cm}{!}{\includegraphics*[0,430][610,715]{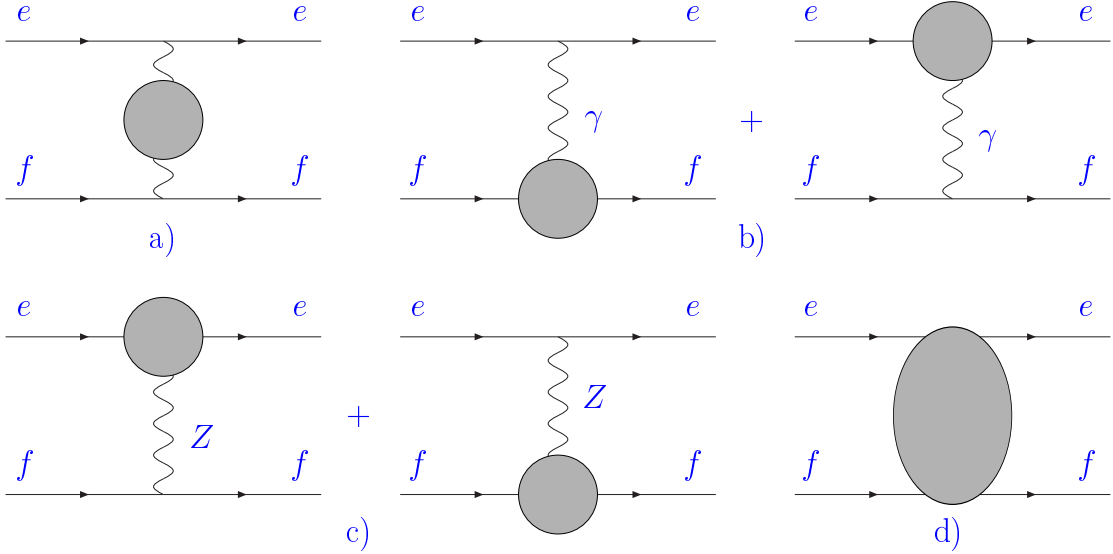}}
\caption{Various types of radiative corrections to parity-violating electron
scattering: $Z^0$ boson self-energy and $Z-\gamma$ mixing (a), anapole
moment contributions (b), vertex corrections (c), and box graphs
(d). External leg corrections are not explicitly shown.}
\label{fig:nc-general}
\end{center}
\end{figure}
The SM values for $\delta\rho_{PV}$, $\delta{\kappa}_{PV}$, and
$\hat\lambda_{iq}$ are well-known\cite{pdg}. When R parity is an exact
symmetry of the MSSM, the presence of virtual SUSY particles will
modify these quantities solely via the loop amplitudes of
Fig.~\ref{fig:nc-general}. The SUSY contributions to ${\rho}_{PV}$
and ${\kappa}_{PV}$ have been worked out in
Ref. \cite{susypv}. These contributions, which are are universal and
can be applied to all low-energy PV processes, can be expressed in
terms of the oblique parameters\cite{susypv}: \beqa
\label{eq:rho-kappa-stu}
\delta\rho_{PV}^{\rm SUSY} & = & {\hat\alpha} T-{\hat\delta}_{VB}^\mu,
\nonumber \\ \delta\kappa_{PV}^{\rm SUSY} & = & \left({{\hat c}^2\over
{\hat c}^2-{\hat s}^2} \right) \left({{\hat\alpha}\over 4{\hat s}^2
{\hat c}^2} S-{\hat \alpha} T +{\hat\delta}_{VB}^\mu \right) \nonumber
\\ &&+{{\hat c}\over {\hat s}}\Bigl[ {{\hat\Pi}_{\gamma Z}(q^2)\over
q^2}- {{\hat\Pi}_{\gamma Z}(M_Z^2)\over M_Z^2}\Bigr]^{\rm SUSY}
+\Bigl({{\hat c}^2\over {\hat c}^2-{\hat s}^2}
\Bigr)\Bigl[-{{\hat\Pi}_{\gamma\gamma}(M_Z^2)\over M_Z^2}
+{\Delta{\hat\alpha}\over \alpha} \Bigr]^{\rm SUSY}.  \eeqa

Note that we have modified the definition of $\delta\kappa^{\rm
susy}_{PV}$ compared to that given in Ref. \cite{susypv}, moving the
anapole contribution into the non-universal correction terms
$\hat\lambda_{iq}$. The quantities ${\hat\Pi}_{VV'}(q^2)$ are the
gauge boson propagator functions, renormalized in the $\overline{\rm
DR}$ scheme\footnote{Quantities renormalized in the $\overline{\rm
DR}$ scheme are indicated by a hat.}, and the oblique parameters $S$,
and $T$ are defined in terms of various combinations of these
quantities as given in Ref. \cite{peskin}.  
The contribution ${\hat\delta}^\mu_{VB}$
denotes SUSY vertex, external leg, and box graph corrections to the
amplitude for muon decay. These corrections must be subtracted from
the semileptonic, neutral current
amplitude since the latter is 
normalized in terms of the Fermi constant obtained from the
muon lifetime. The quantity
$\Delta{\hat\alpha}$ is the SUSY contribution to the difference
between the fine structure constant and the electromagnetic coupling
renormalized at $\mu=M_Z$:
$\Delta{\hat\alpha}=[{\hat\alpha}(M_Z)-\alpha]^{\rm SUSY}$. Note that
gauge invariance implies that ${\hat\Pi}_{Z\gamma}(q^2)/q^2$ is finite
as $q^2\to 0$.

For the ${\hat\lambda}_{1,2q}$ one has
\begin{eqnarray}
{\hat\lambda}_{1q}^{\rm SUSY}&=& g_Z^{eA}(g_{Z}^{qV}\delta
Z_V^q-g_{Z}^{qA}\delta Z_A^q+ G_V^q) +g_{Z}^{qV}(g_Z^{eA}\delta
Z_V^e-g_Z^{eV}\delta Z_A^e+ G_A^e) \nonumber \\ && +{\hat \delta}_{\rm
box}^{eAqV} - 16 Q_q \hat{c}^2\hat{s}^2 F_{A,e} \\ {\rm
\lambda}_{2q}^{\rm SUSY}&=& g_Z^{qA}(g_{Z}^{eV}\delta
Z_V^e-g_{Z}^{eA}\delta Z_A^e+ G_V^e) +g_{Z}^{eV}(g_Z^{qA}\delta
Z_V^q-g_Z^{qV}\delta Z_A^q+ G_A^q) \nonumber \\ && +{\hat \delta}_{\rm
box}^{eVqA} - 16 Q_e \hat{c}^2\hat{s}^2 F_{A,q}
\end{eqnarray}
where the various couplings, counterterms as well as the vertex
corrections are defined in Appendix A of Ref. \cite{susypv}.
In particular, the $g_Z^{fV}$ ($g_Z^{fA}$) is the vector (axial vector)
coupling of fermion $f$ to the $Z^0$;  $\delta Z^q_V$ ($\delta Z^q_A$)
is $1/2\times$ the sum (difference) of right- and left-handed fermion
wave function renormalization constants; $G_V^f$ ($G_A^f$) give the loop
contributions to the vector (axial vector) $Zff$ vertex; 
 and in the notation introduced in Eq. (C14) of Ref. \cite{susypv} \beqa {\hat
\delta}_{\rm box}^{eAqV}&=&-2\left(-A^{ef}+B^{ef}-C^{ef}+D^{ef}\right)
\nonumber \\ {\hat \delta}_{\rm
box}^{eVqA}&=&-2\left(-A^{ef}+B^{ef}+C^{ef}-D^{ef}\right) \eeqa
are the box diagram contributions to the $A(e)\times V(q)$ and $V(e)\times
A(q)$ amplitudes, respectively.

\begin{figure}[ht]
\begin{center}
\resizebox{16. cm}{!}{\includegraphics*[0,580][600,720]{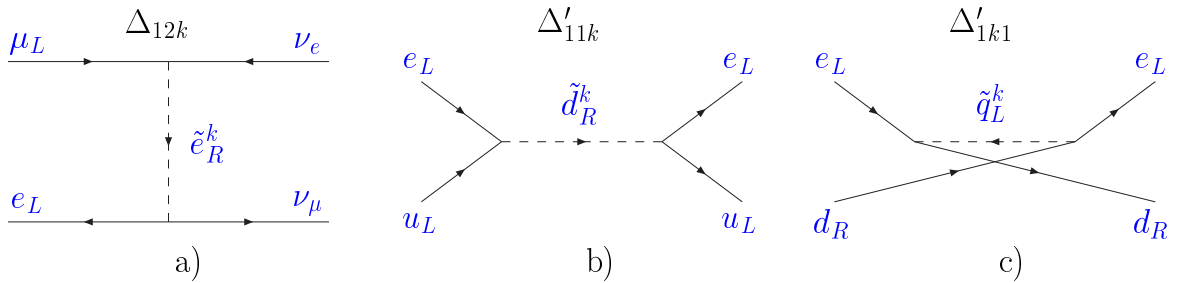}}
\caption{
Tree-level RPV contributions to (a)  muon decay, (b)
electron-$u$-quark scattering, and (c) 
electron-$d$-quark scattering. The quantities
$\Delta_{ijk}^{(\prime)}$ are defined in Eq. (\ref{eq:deltas}).  }
\label{fig:rpv-graphs}
\end{center}
\end{figure}
New tree-level SUSY contributions to the $C_{iq}$ can be generated
when the R parity quantum number $P_R=(-1)^{3(B-L)+2S}$ is not
conserved, with $B$, $L$, and $S$ denoting, respectively, the baryon
number, lepton number, and spin of a given particle. In order to avoid
unacceptably large contributions to the proton decay rate, we do not
allow for any $\Delta B\not= 0$, $\Delta L=0$  R parity violating (RPV)
interactions. Those which violate $L$ are parameterized by the
couplings $\lambda_{ijk}$ and $\lambda^\prime_{ijk}$ appearing in the
superpotential, where the former involve purely leptonic interactions
and the latter corresponding to semileptonic effects. The subscripts
denote the generations of the three particles in the interaction. The
corresponding contributions to the $C_{iq}$ arise from the Feynman
diagrams of Fig.~\ref{fig:rpv-graphs} and involve the positive,
semi-definite, dimensionless quantities:
\begin{equation}
\Delta_{ijk}({\tilde f}) = {|\lambda_{ijk}|^2\over 4\sqrt{2} G_\mu
M^2_{\tilde f_k}} \ \ \ ,
\label{eq:deltas}
\end{equation}
where ${\tilde f}_k$ is the scalar superpartner of fermion $f_k$ and where
an analogous expression applies for the semi-leptonic corrections,
$\Delta_{ijk}^\prime({\tilde f})$.  Note that one must consider both
the purely leptonic and semileptonic corrections, as the former modify
the relative normalization of neutral and charged current amplitudes
as well as the predicted value of the weak mixing angle\cite{mrm00}.

In terms of the $\Delta_{ijk}({\tilde f})$ and
$\Delta_{ijk}^\prime({\tilde f})$, one has the following shifts in the
$C_{iq}$:
\begin{eqnarray}
\Delta C_{1u}^{\rm RPV} & = & -[C_{1u}-\frac{4}{3}\lambda_x
]\Delta_{12k}({\tilde e}^k_R)-\Delta^\prime_{11k}({\tilde d}^k_R) \\
\Delta C_{1d}^{\rm RPV} & = & -[C_{1d}+\frac{2}{3}\lambda_x
]\Delta_{12k}({\tilde e}^k_R)+\Delta^\prime_{1k1}({\tilde q}^k_L)\\
\Delta C_{2u}^{\rm RPV} & = & -[C_{2u}-2\lambda_x
]\Delta_{12k}({\tilde e}^k_R)-\Delta^\prime_{11k}({\tilde d}^k_R)\\
\Delta C_{2d}^{\rm RPV} & = & -[C_{2d}+2\lambda_x
]\Delta_{12k}({\tilde e}^k_R)-\Delta^\prime_{1k1}({\tilde q}^k_L)\ \ \
,
\end{eqnarray}
where
\begin{equation}
\lambda_x = {{\hat s}^2(1-{\hat s}^2)\over 1-2{\hat s}^2}{1\over
1-\Delta {\hat r}^{SM}}\approx 0.35
\end{equation}
with $\Delta {\hat r}^{SM}$ being a radiative correction to the
$\alpha$-$G_\mu$-$M_Z$-$\sstw$ relation in the SM:
\begin{equation}
{\hat{s}}^2{\hat{c}}^2=\frac{\pi \alpha}{\sqrt{2} M_Z^2 G_\mu
(1-\Delta{\hat r}^{SM})}.
\end{equation}

We first analyze the SUSY corrections to the $C_{iq}$ and $\alred$
assuming conservation of R parity. To do so, we randomly generate
$\sim$ 5000 different sets of soft SUSY-breaking parameters, chosen
from a flat distribution in the SUSY-breaking mass parameters and a
logarithmic distribution in $\tan\beta$. The ranges chosen for the
parameters are given in Table \ref{tab:param}.
\begin{table}
\begin{center}
\begin{minipage}[t]{16.5 cm}
\caption[]{Range of MSSM parameters chosen for computation of SUSY 
radiative corrections. Here $\tilde M$ denotes any of the gaugino masses, 
sfermion masses or the $\mu$ parameter, while  $A_f$ denotes the triscalar 
couplings that enter the off diagonal term in the mass-squared matrix for 
scalar fermion
$\tilde f$.}
\label{tab:param}
\vspace*{4pt}
\end{minipage}
\begin{tabular}{|c|c|c|c|c|c|c|c|c|}
\hline &\\[-8pt] parameter & range \\[4pt] \hline &\\[-8pt]
$\tan\beta$ & 1.4 $\rightarrow$ 60 \\ $\tilde M$ & (50 $\rightarrow$
1000) GeV \\ $A_f$ & ($-1000$ $\rightarrow$ $1000$) GeV$$\\ [-8pt] &
\\ \hline
\end{tabular}
\end{center}
\end{table}     
The parameters varied include the gaugino masses, Higgs bilinear mass
parameter $\mu$, $\tan\beta$, sfermion masses, and the triscalar
couplings $A_f$. The latter enter the off-diagonal entries in the
mass-squared matrix for scalar fermion $\tilde f$:
\begin{equation}
(M_{\tilde f}^2)^i_{LR} = \cases{ m_f(A_f-\mu\cot\beta), & $Q_f > 0$
\cr m_f(A_f-\mu\tan\beta), & $Q_f <0$ \cr }\ \ \ .
\end{equation}
The lower limit on $\tan\beta$ is derived from an analysis of
electroweak symmetry-breaking and direct bounds on the lightest SUSY
Higgs searches from LEP. The lower bounds on the gaugino and sfermion mass
parameters correspond roughly to direct search lower bounds on SUSY
masses, while the limits on $A_f$ is taken from naturalness bound.

For each parameter set, we compute the SUSY radiative corrections to
 the $C_{iq}$.  In order to incorporate what is known
 phenomenologically, we impose several constraints on the acceptable
 parameter sets. First, to avoid unacceptably large flavor-changing
 neutral currents, we do not allow for generation mixing\footnote{In
 practice, we have also taken the CKM matrix for quarks to be
 diagonal. For the observables of interest here, this approximation
 introduces negligible error.}.  Second, we retain only those
 parameter sets that give values for $S$ and $T$ consistent with
 current constraints on these parameters. As discussed in
 Ref. \cite{susypv}, neglecting the non-oblique corrections to the
 $Z$-pole precision observables entails some lack of self-consistency,
 but does not distort the qualitative conclusions.

\begin{figure}[ht]
\hspace{0.00in}
\begin{center}
\resizebox{10.cm}{!}{\includegraphics*[30,190][520,600]{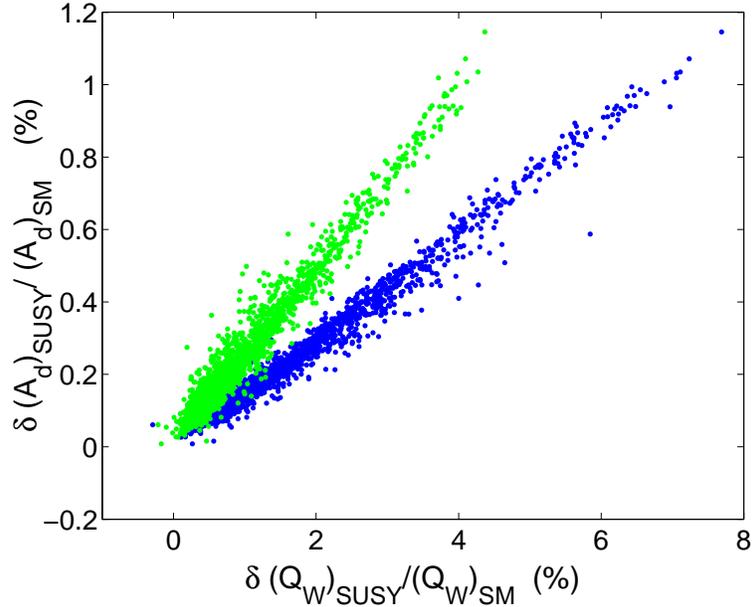}}
\caption{SUSY loop correction to the relative shift in $\alred(y=1)$ 
vs. the relative shifts in the electron (dark dots) and proton weak 
charges (light dots).}
\label{fig:loop}
\end{center}
\end{figure}

In presenting our results, we follow the spirit of our previous work
and plot in Fig.~\ref{fig:loop} the relative shift in $\alred$ vs. the
relative shifts in the electron and proton weak charges. The weak
charge, $Q_W^f$, of particle $f$  characterizes the effective $A(e)\times V(f)$
neutral current interaction:
\begin{equation}
{\cal L}^{PV,\ \rm eff}_{ef} = - {G_\mu\over 2\sqrt{2}}Q_W^f A(e)\cdot 
V(f)\ \ \ ,
\end{equation}
which can be measured in PV $ee$ and elastic PV $ep$ scattering experiments.
In terms of the $C_{iq}$ one has for the proton $Q_W^p=-2(2C_{1u}+C_{1d})$.
In the SM, the weak charges of the proton and electron are suppressed, making
them relatively transparent to the effects of new physics. 

As with the SUSY loop effects in $Q_W^e$ and $Q_W^p$, the shifts induced in
$\alred$ are highly correlated with those in the electron and proton
weak charges. In contrast, the loop SUSY contributions to the weak
charge of cesium are negligible as a result of
cancellations between up and down quark contributions\cite{susypv}. 
Thus, the present agreement of atomic PV
with the SM does not preclude substantial SUSY loop contributions to
$\alred$. Indeed, the magnitude of the corrections can be as much as
one percent of the SM asymmetry for sufficiently light SUSY mass
parameters or large $\tan\beta$.  From a practical standpoint, one
would need to perform a measurement with precision better than $\sim
0.5\%$ in order to be sensitive to the effects of SUSY radiative
corrections.  A recent Letter of Intent for a measurement at SLAC
proposed a combined statistical and systematic error of 0.8 \%
\cite{slacloi}, suggesting that, with further refinements, a sensitive
probe of SUSY loop contributions may be feasible in the future.

\begin{figure}[ht]
\hspace{0.00in}
\begin{center}
\resizebox{10.cm}{!}{\includegraphics*[30,180][560,600]{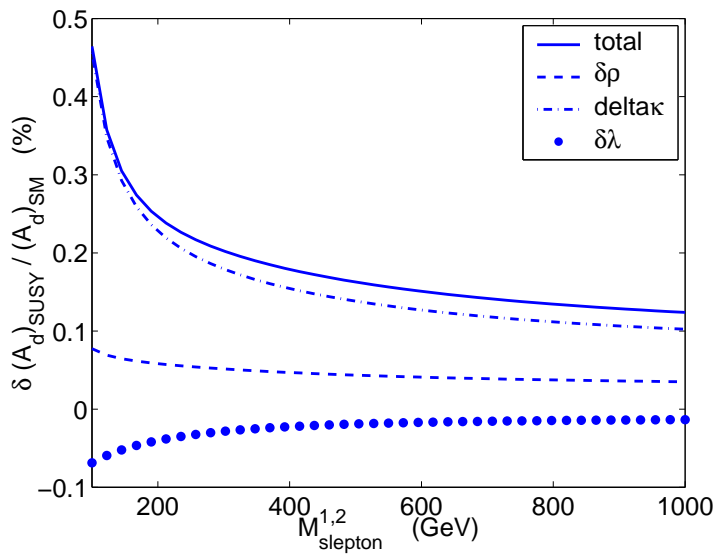}}
\caption{
SUSY contributions to $\alred(y=1)$ from $\delta \rho_{PV}$ (dashed line), 
$\delta
\kappa_{PV}$ (dash-dotted line) and $\hat\lambda$ (dotted line).  The solid
line is the sum of all the contributions to $\alred$.  The $x$-axis is
the common first and second generation slepton mass.  The other MSSM
parameters are chosen to be $\tan\beta=10$, $2M_1=M_2=\mu=200$ GeV.
The mass for the squarks and third generation slepton are taken to be
1000 GeV.  Note that neutralinos and charginos are light, which leads
to a non-decoupling effect in $\alred$ even when squarks and
sleptons are heavy.  }
\label{fig:detail}
\end{center}
\end{figure}

We observe, however, that the strong correlation of loop effects
between $\alred$ and the weak charges makes the combination of all
three measurements a potentially interesting probe of SUSY radiative
corrections. The reason for this correlation is that the contributions
to $\delta{\kappa}_{PV}$ -- and, thus, the shift in the effective weak
mixing angle -- dominate the loop-induced corrections to the asymmetry
(see Fig.~\ref{fig:detail}). A similar statement applies to the
electron and proton weak charges. Given the precision anticipated for
the measurements of $Q_W^e$ and $Q_W^p$ as well as the precision
proposed in \cite{slacloi} for $\alred$, the combination of all three
measurements would be nearly a factor of two more sensitive to SUSY
loop-induced shifts in $\sin^2\theta_W^{\rm eff}=\kappa_{PV} \hat{s}^2$ 
than of the measurements individually.

In Fig.~\ref{fig:qwe_qwp}, we illustrate the sensitivity of $\alred$
to the effects of RPV interactions. Here, we plot the relative shifts
in $\alred$ vs. those in $Q_W^e$ and $Q_W^p$ due to both SUSY loops as
well as RPV interactions. The interior of the truncated ellipse gives
the 95\% C.L. region from RPV effects 
allowed by other precision electroweak data. The
latter include $\mu$-decay, $\beta$-decay, $\pi$-decay, cesium atomic
PV, and $M_W$. The constraints from $Z^0$-pole observables are fairly
insensitive to these RPV effects since they generate off-resonance
tree-level contributions to $e^+e^-$ annihilation amplitudes. The
truncation of the elliptical region results from the requirement that
the corrections $\Delta_{ijk}$ and $\Delta^\prime_{ijk}$ be
non-negative.

\begin{figure}[ht]
\hspace{0.00in}
\begin{center}
\resizebox{8.cm}{!}{\includegraphics*[30,200][520,600]{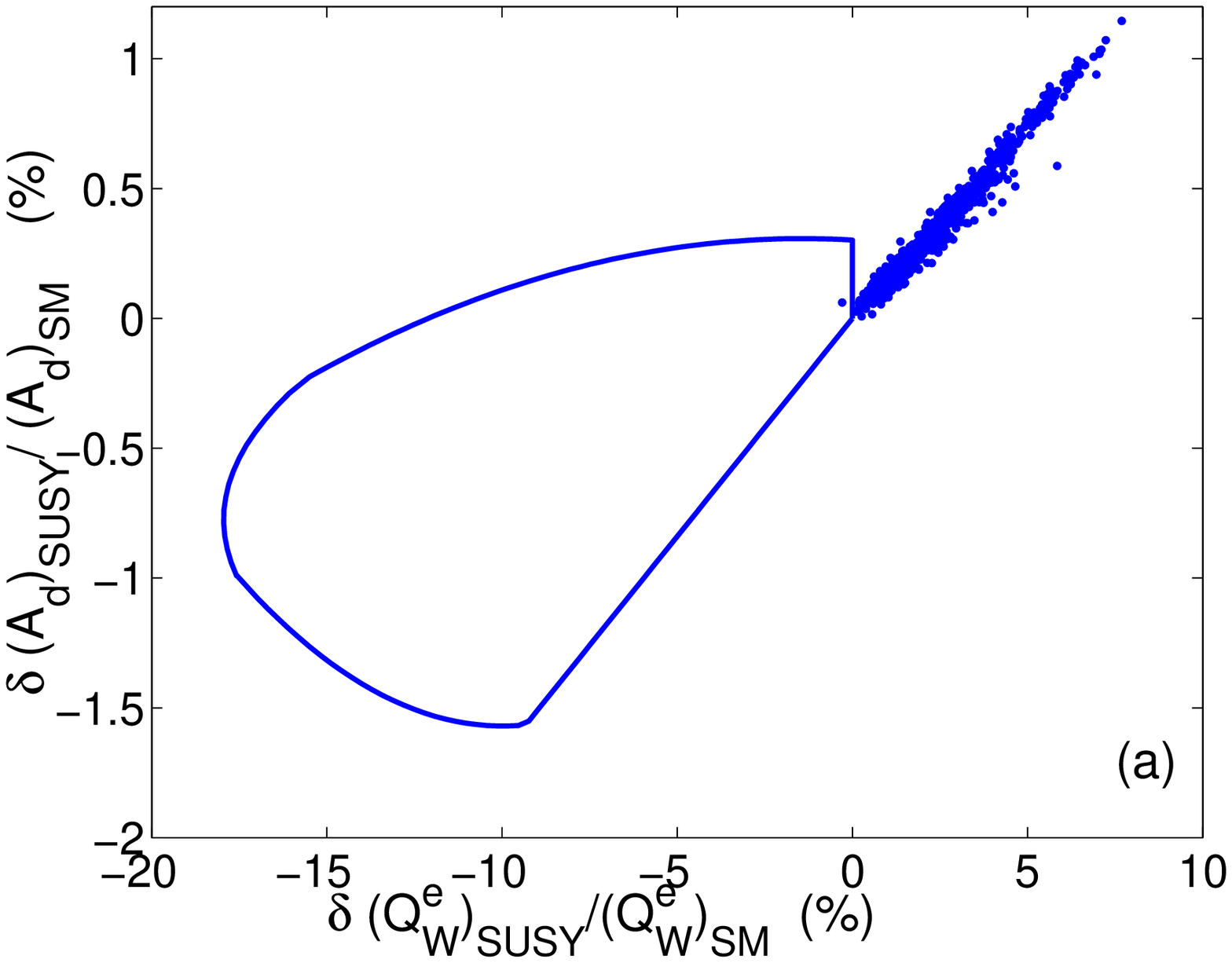}}
\resizebox{8.cm}{!}{\includegraphics*[30,200][520,600]{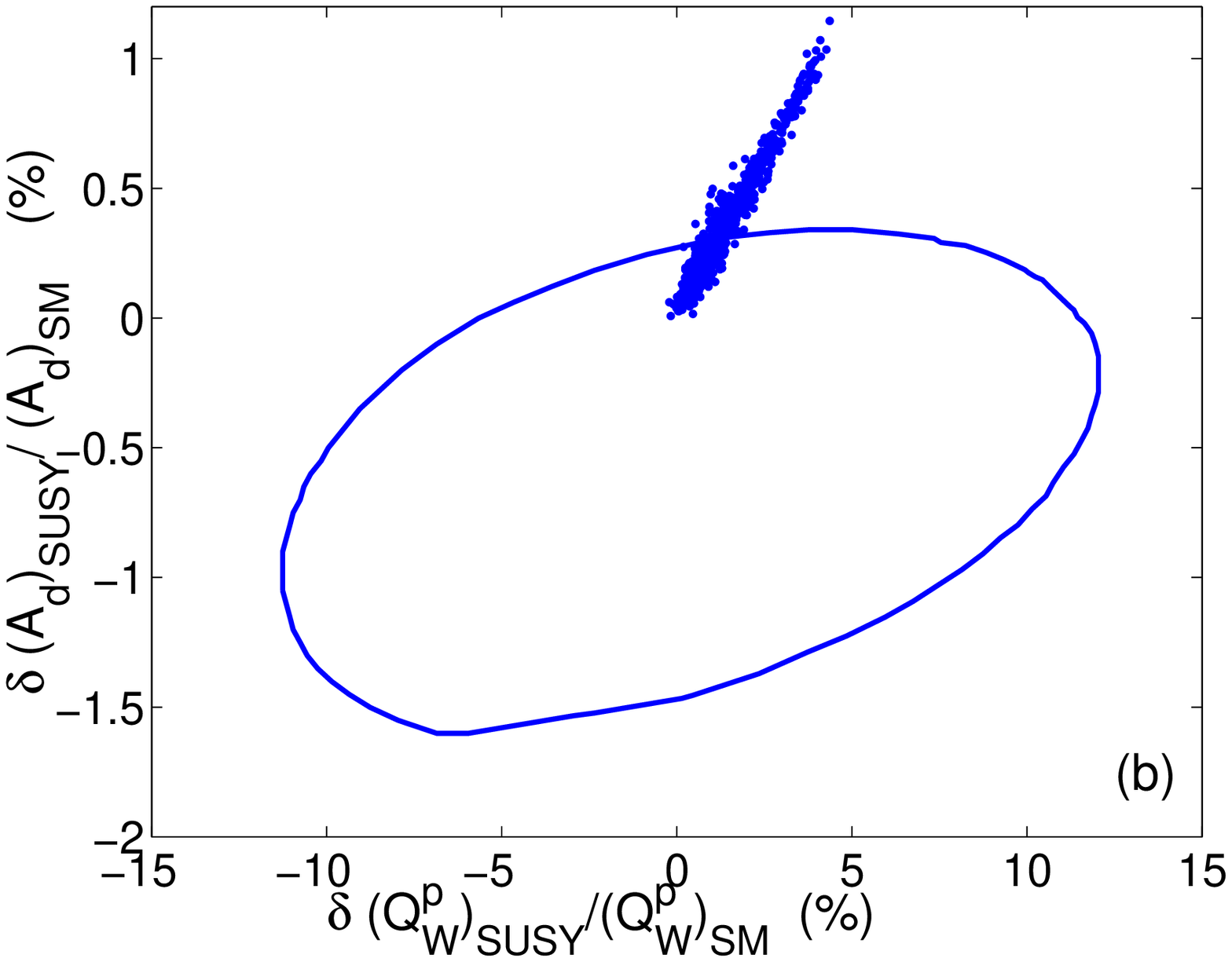}}
\caption{95 $\%$ CL allowed region for PRV contribution to $\alred(y=1)$ 
vs. electron weak charge (a) and proton weak charge (b).  
The dots indicate the SUSY loop corrections. }
\label{fig:qwe_qwp}
\end{center}
\end{figure}

As illustrated in by Fig.~\ref{fig:qwe_qwp} (a), the presence of RPV
effects would induce negative relative shifts in both $\alred$ and
$Q_W^e$, whereas the relative sign of the loop corrections is positive
in both cases.  Moreover, the maximum correction to $\alred$ would be
$-1.5\%$, corresponding to about 2$\sigma$ for the precision proposed
in Ref. \cite{slacloi}. The corresponding comparison of the
corrections to $\alred$ and $Q_W^p$ are shown in
Fig.~\ref{fig:qwe_qwp}(b). Note that in this case, a sizable positive
shift in $Q_W^p$ [up to $3\sigma$ for the proposed $\alr(ep)$
measurement] due to RPV contributions could correspond to a tiny
effect on $\alred$ whereas a substantial negative shift in the proton
weak charge could also occur in tandem with a substantial negative
correction to $\alred$. On the other hand, even a result for $Q_W^p$
consistent with the SM would not rule out a sizable effect on
$\alred$.

In summary, we conclude that a combination of PV electron scattering
measurements at kinematics below the $Z^0$ pole may provide an
interesting probe of SUSY. Given the correlations between the SUSY
loop and RPV effects among the different observables, a comparison of
PV elastic $ee$, elastic $ep$ and deep-inelastic $eD$ scattering 
would provide substantially more information than any one
alone. In this context, the addition of an $\alred$ measurement would
provide a useful complement to the PV $ee$ and elastic $ep$
measurements, assuming it can be performed with $\sim 0.5\%$ precision
or better. More generally, a comprehensive program of PV electron
scattering experiments -- when combined with the results of atomic PV
deep-inelastic $\nu$-nucleus scattering -- could help determine which
versions of SUSY remain the most viable.

\begin{acknowledgments}
We thank R. Arnold, P. Bosted, R. Holt, and P. Reimer for useful
discussions.  This work is supported in part under U.S. Department of
Energy contract \#DE-FG03-02ER41215 (A.K. and M.J.R.-M.),
\#DE-FG03-00ER41132 (M.J.R.-M.), and \#DE-FG03-92-ER-40701 (S.S.).
A.K. and M.J.R.-M. are supported by the National Science Foundation
under award PHY00-71856.  S.S. is supported by the John A. McCone
Fellowship.
\end{acknowledgments}

\end{document}